\documentclass[twocolumn,showpacs,preprintnumbers,amsmath,amssymb]{revtex4}
\usepackage{amsmath,amssymb,graphics,epsfig,subfigure}
\usepackage{color}

\begin{document}

\thispagestyle{empty}

\begin{center}

\title{Topological Charge and Black Hole Photon Spheres}

\date{\today}
\author{Shao-Wen Wei \footnote{E-mail: weishw@lzu.edu.cn}}

\affiliation{Institute of Theoretical Physics $\&$ Research Center of Gravitation, Lanzhou University, Lanzhou 730000, People's Republic of China,\\
and Joint Research Center for Physics, Lanzhou University and Qinghai Normal University, Lanzhou 730000 and Xining 810000, China}

\begin{abstract}
Black hole photon spheres or light rings are closely linked to astronomical phenomena, such as the gravitational waves and the shadow in spherically symmetric or axi-symmetric spacetime. In Cunha and Herdeiro [Phys. Rev. Lett. 124, 181101 (2020)], a topological argument was applied for the four dimensional stationary, axi-symmetric, asymptotically flat black hole, and the result indicates that at least, there exists one standard light ring outside the black hole horizon for each rotation sense. Inspired by it, in this paper, we would like to consider a similar issue for a nonrotating, static, spherically symmetric black hole not only with asymptotically flat behavior, but also with AdS and dS behaviors. Following Duan's topological current $\phi$-mapping theory, the topological current and charge for the photon spheres are introduced. The topological current is nonzero only at the zero point of the vector field determining the location of the photon sphere. So each photon sphere can be assigned a topological charge. Considering the full exterior region, we find the total topological charge always equals -1. This result confirms that there exists at least one standard photon sphere outside of the black hole not only in asymptotically flat spacetime, but also in asymptotically AdS and dS spacetime. Then we apply the study to the dyonic black holes. We observe that even when more photon spheres are included, the total topological charge stays unchanged. Moreover, for a naked singularity, it has a vanishing topological charge, indicating that the black hole and naked singularity are in different topological classes. It is expected that this novel topological argument could provide an insightful idea on the study of the black hole photon spheres or light rings, and further cast new light on the black hole astronomical phenomena.
\end{abstract}


\maketitle
\end{center}

\section{Introduction}
\label{secIntroduction}

The observations of gravitational waves \cite{Abbott} and black hole shadow image \cite{Akiyama1,Akiyama2,Akiyama3} mark a new era of astrophysical observations. These provide powerful tests to uncover the spacetime structure near a black hole horizon.

In the ringdown stage of a black hole merge, the radiating gravitational waves can be understood by the quasinormal modes. It is also well known that in the eikonal limit, the quasinormal modes are linked to the photon spheres (PSs) or light rings (LRs) of the nonrotating or rotating black holes (for recent progress, see Refs. \cite{Cardoso,Glampedakis,Kokkotas,Franzin}). On the other hand, the formation of a black hole shadow mainly depends on the existence of a PS or LR rather the horizon. Moreover, the relativistic images of a light source around a compact object are closely dependent on a PS and LR. Different relations between these observable phenomena were studied in Refs. \cite{Decanini,Stefanov,Wei2011,Wei2014,Herdeiro,Claudel}. All the results confirm that a PS and LR play an extremely important role in astronomical observation. Therefore, it is valuable to investigate the characteristic properties of PS or LR for a certain spacetime.

In Ref. \cite{Cunhaa}, Cunha, Berti, and Herdeiro considered the LR stability for ultra-compact objects. Based on the Brouwer degree of a continuous map, they found that the LRs of the compact objects always come in pairs. This introduces a novel scheme to investigate the LRs by using a topological argument, while ignoring the specific field equations. This result is proved to be generally true, however, there is an important exception when the degenerate light rings are present \cite{Hoda,Hodb}. Recently, Cunha and Herdeiro \cite{Cunhab} put forward a big step, and generalized the study to a four dimensional stationary, axi-symmetric, asymptotically flat black hole spacetime. They proved that there exists, at least, one standard LR outside the black hole horizon for each rotation sense by calculating the winding number of the vector field defined by an effective potentials on the orthogonal ($r$, $\theta$)-space. In addition, following their topological argument, a horizonless ultra-compact object, such as the boson star, will show an even number of non-degenerate LRs. A valuable issue concerning the topological argument is to consider the non-rotating black holes. In addition to the asymptotically flat case \cite{Cunhaa}, the study can also be extended to these black holes with AdS and dS behaviors. This can help us to understand the topological property of PSs, and to uncover the influence of black hole spin on the PS topology.

From an another point of view, we in Ref. \cite{Wei2019} also introduced an interesting topological approach to investigate the black hole shadow caused by the existence of the LRs or PSs. Each shadow shape is endowed with a topological covariant quantity. For a black hole shadow, it equals one, while it deviates one for a naked singularity. For a multiple disconnected shadow pattern, it can produce the number of the shadow. So through the topological quantity, one can distinguish different spacetimes. This also provides us a distinctive approach to study the black hole shadow from the topological argument.

In Refs. \cite{Cunhaa,Cunhab}, the authors calculated the winding number, a important topological quantity, of the vector field around the zero point of the vector field, where a LR is located. Thus the topological properties of the spacetime were revealed by that topological quantity. Similarly, there is a famous theory known as Duan's topological current $\phi$-mapping theory \cite{Duana,Duanb}, which starts the topological current and then the corresponding topological charge related to the winding number is given. This powerful approach has shown insightful ideas into different physical systems, such as gauge theories, superconduction, monopoles, magnetic skyrmions, knots, cosmological strings, quantum Hall effect, and so on. Thus, it provides us a natural and effective tool to investigate the topological structure of PSs. In this paper, we will follow Refs. \cite{Duana,Duanb} to study the topology of PSs in a general static, spherically symmetric black hole with asymptotically flat, AdS, dS behaviors of boundary.

The present paper is organized as follows. In Sec. \ref{bhaps}, we give a brief introduction of PS for a non-rotating black hole. Then we follow Refs. \cite{Cunhaa,Cunhab} and introduce the vector field through a regular potential function outside a black hole horizon on the ($r$, $\theta$) plane. Then based on the vector field, we in Sec. \ref{tcac} present the topological current and charge. In particular, following the topological current $\phi$-mapping theory, we express the topological current in terms of a $\delta$ function. The inner structure of the topological charge is also investigated. In Sec. \ref{bhatc}, we compute the topological charge for a black hole with asymptotically flat, AdS, and dS behaviors of boundary. All these black holes are found to admit a minus one topological charge indicating the existence of at least one standard PS. Next, in Sec. \ref{dbh}, as a specific example, we calculate the topological charge for the dyonic black holes. Three different cases are examined in details. After that, we briefly discuss the annihilation of PSs in Sec. \ref{psa}. Finally, we summarize and discuss our results in Sec. \ref{Conclusion}.

\section{Black holes and photon spheres}
\label{bhaps}

In this paper, we only consider the static, spherically symmetric black hole. The black hole solution is assumed to be in the following form:
\begin{equation}
 ds^2=-f(r)dt^2+\frac{1}{g(r)}dr^{2}+h(r)(d\theta^2+\sin^2\theta d\varphi^2).\label{metric}
\end{equation}
Generally, the radius $r_{\texttt{h}}$ of the black hole horizon is the largest root of $g(r_{\texttt{h}})$=0 or $f(r_{\texttt{h}})$=0. On the other hand, by solving the null geodesics, one can obtain the radial motion on the equatorial plane
\begin{equation}
 \dot{r}^2+V_{\texttt{eff}}=0,
\end{equation}
where the effective potential is given by
\begin{equation}
 V_{\texttt{eff}}=g(r)\left(\frac{L^2}{h(r)}-\frac{E^2}{f(r)}\right).
\end{equation}
Here $E$ and $L$ are the energy and angular momentum of photon, which are related with the Killing vector fields $\partial_t$ and $\partial_{\phi}$, respectively. Since this solution is spherically symmetric, there exists a PS at $r_{\texttt{ps}}$ determined by
\begin{equation}
 V_{\texttt{eff}}=0, \quad \partial_{r}V_{\texttt{eff}}=0.
\end{equation}
Solving them, we find that the radius of the PS satisfies the following equation:
\begin{equation}
 \left(\frac{f(r)}{h(r)}\right)_{r=r_{\texttt{ps}}}'=0,\label{pphoton}
\end{equation}
where the prime indicates the derivative with respect to $r$. Moreover, $\partial_{r,r}V_{\texttt{eff}}(r_{\texttt{ps}})$ $<$($>$)0 indicates the PS is unstable (stable). Carrying out the derivative, (\ref{pphoton}) reduces to
\begin{equation}
 f(r)h(r)'-f(r)'h(r)=0.\label{pphoton2}
\end{equation}
It is worth noting that at the horizon where $f(r_{\texttt{h}})$=0, the first term vanishes, while the second term is generally nonzero. So the locations of $r_{\texttt{ps}}$ and $r_{{\texttt{h}}}$ are different. However, when a black hole has more than one horizon, there will be the extremal black hole case, where two horizons coincide. Or for the extremal black hole case, we both have both $f(r_{\texttt{h}})=0$ and $f(r_{\texttt{h}})'=0$. Significantly, condition (\ref{pphoton2}) is satisfied. So the PS and the extremal black hole horizon naturally coincide for the extremal black hole.

In order to study the topological property of the PS, we introduce the everywhere regular potential function \cite{Cunhaa}
\begin{equation}
 H(r, \theta)=
 \sqrt{\frac{-g_{tt}}{g_{\varphi\varphi}}}
 =\frac{1}{\sin\theta}\left(\frac{f(r)}{h(r)}\right)^{\frac{1}{2}}.
\end{equation}
Obviously, the radius of the PS locates at the root of $\partial_rH$=0. Similar to Ref. \cite{Cunhab}, we can introduce a vector field $\phi$=($\phi^r$, $\phi^\theta$) \footnote{In Ref. \cite{Cunhab}, the notation of the vector field is $v$. Here in order to match the Duan's topological current $\phi$-mapping theory, we label it with $\phi$.}
\begin{equation}
 \phi^r=\frac{\partial_rH}{\sqrt{g_{rr}}}=\sqrt{g(r)}\partial_rH,\quad
 \phi^\theta=\frac{\partial_\theta H}{\sqrt{g_{\theta\theta}}}
  =\frac{\partial_\theta H}{\sqrt{h(r)}}.\label{vectorfield}
\end{equation}
Although the circular photon orbit for a spherically symmetric black hole is a PS, which is independent of the coordinate $\theta$, here we aim to investigate the topological property of the circular photon orbit, so we preserve $\theta$ in our discussions. Note that the vector can also be reformulated as
\begin{equation}
 \phi=||\phi||e^{i\Theta},\label{pp9}
\end{equation}
where $||\phi||=\sqrt{\phi^a\phi^a}$.
However, in terms of $\phi$, a PS occurs at $\phi$=0. This implies that $\phi$ in (\ref{pp9}) is not well defined for the PS, so we treat the vector as $\phi=\phi^r+i\phi^\theta$. The normalized vectors are defined as
\begin{equation}
 n^a=\frac{\phi^a}{||\phi||}, \quad a=1,2,\label{nonon}
\end{equation}
with $\phi^1=\phi^r$ and $\phi^2=\phi^\theta$.

\section{Topological current and charge}
\label{tcac}

In this section, by treating the PSs as the defects located at the zero points of $\phi$, we will study their topological current and charge following Duan's $\phi$-mapping topological current theory.

At first we define a superpotential
\begin{equation}
 V^{\mu\nu}=\frac{1}{2\pi}\epsilon^{\mu\nu\rho}\epsilon_{ab}n^{a}\partial_{\rho}n^{b},
 \quad\mu,\nu,\rho=0,1,2.
\end{equation}
Here $x^{\mu}$=($t$, $r$, $\theta$). Note that one can reformulate the coordinate $t$ with other black hole parameters as we will show in what follows. It is clear that the superpotential is an antisymmetric tensor $V^{\mu\nu}=-V^{\nu\mu}$. Employing the superpotential, we introduce a topological current
\begin{equation}
 j^{\mu}=\partial_{\nu}V^{\mu\nu}
  =\frac{1}{2\pi}\epsilon^{\mu\nu\rho}\epsilon_{ab}\partial_{\nu}n^{a}\partial_{\rho}n^{b}.
  \label{curr}
\end{equation}
It is easy to find that this topological current satisfies
\begin{equation}
 \partial_{\mu}j^{\mu}=0.
\end{equation}
The component $j^0$ is the charge density. Integrating it, we will obtain the topological charge at given $\Sigma$,
\begin{equation}
 Q=\int_\Sigma j^{0}d^{2}x.
\end{equation}
In the next, we aim to uncover the characteristic property of the topological current $j^{\mu}$. Inserting (\ref{nonon}) into (\ref{curr}), we have
\begin{equation}
 j^{\mu}=\frac{1}{2\pi}\epsilon^{\mu\nu\rho}\epsilon_{ab}\frac{\partial}{\partial\phi^{c}}
 \left(\frac{\phi^a}{||\phi||^2}\right)\partial_{\nu}\phi^{c}\partial_{\rho}\phi^{b}.
\end{equation}
Note that $\frac{\partial \ln||\phi||}{\partial\phi^a}=\frac{\phi^a}{||\phi||^2}$,
we can express the topological current as
\begin{equation}
 j^{\mu}=\frac{1}{2\pi}\epsilon^{\mu\nu\rho}\epsilon_{ab}
 \left(\frac{\partial}{\partial\phi^{c}}\frac{\partial}{\partial\phi^{a}}\ln||\phi||\right)
 \partial_{\nu}\phi^{c}\partial_{\rho}\phi^{b}.
\end{equation}
In terms of the Jacobi tensor
\begin{equation}
 \epsilon^{ab}J^{\mu}\left(\frac{\phi}{x}\right)=\epsilon^{\mu\nu\rho}
 \partial_{\nu}\phi^a\partial_{\rho}\phi^b,\label{jaco}
\end{equation}
we get
\begin{equation}
 j^{\mu}=\frac{1}{2\pi}\left(\Delta_{\phi^a}\ln||\phi||\right)J^{\mu}\left(\frac{\phi}{x}\right),
\end{equation}
where $\Delta_{\phi^a}=\frac{\partial}{\partial\phi^a}\frac{\partial}{\partial\phi^a}$. Using the two-dimensional Laplacian Green function in $\phi$-mapping space
\begin{equation}
 \Delta_{\phi^a}\ln||\phi||=2\pi\delta(\phi),
\end{equation}
we have the topological current
\begin{equation}
 j^{\mu}=\delta^{2}(\phi)J^{\mu}\left(\frac{\phi}{x}\right).\label{juu}
\end{equation}
From this expression, we are clear that $j^{\mu}$ is only nonzero at the zero points of $\phi^{a}$, i.e., $\phi^a(x^i, t)=0$. According to the implicit function theorem \cite{Goursat}, when the Jacobi determinant $J^{0}\left(\frac{\phi}{x}\right)\neq0$, one has $(\partial_{\mu}\phi^a)dx^{\mu}=0$. Then from the Jacobi tensor (\ref{jaco}), it is easy to obtain
\begin{equation}
 \epsilon_{\mu\nu\rho}J^{\mu}\left(\frac{\phi}{x}\right)dx^{\nu}=0.
\end{equation}
Multiplying it by $\epsilon^{\lambda\sigma\rho}$, one can arrive at
\begin{equation}
 \frac{dx^{\mu}}{J^{\mu}\left(\frac{\phi}{x}\right)}=\frac{dx^{\nu}}{J^{\nu}\left(\frac{\phi}{x}\right)}.
\end{equation}
After a simple calculation, we have
\begin{equation}
 u^i=\frac{dx^i}{dt}=\frac{J^{i}\left(\frac{\phi}{x}\right)}{J^{0}\left(\frac{\phi}{x}\right)}.\label{uxt}
\end{equation}
Then from (\ref{juu}), the components of $j^{\mu}$ can be expressed in the following form:
\begin{eqnarray}
 j^{i}&=&\delta^{2}(\phi)J^{0}\left(\frac{\phi}{x}\right)u^i,\\
 j^{0}&=&\delta^{2}(\phi)J^{0}\left(\frac{\phi}{x}\right).
\end{eqnarray}
Therefore, the topological charge reads
\begin{eqnarray}
 Q=\int_{\Sigma}\delta^{2}(\phi)J^{0}\left(\frac{\phi}{x}\right)d^2x.\label{qcharge}
\end{eqnarray}
Due to the $\delta$ function, the charge is only nonzero at the zero point of $\phi$, where the PS locates, so we can assign each PS with a topological charge $Q$. When $\Sigma$ covers a single zero point, a detailed study shows that the charge $Q$ exactly equals the winding number. If $\Sigma$ covers several zero points, $Q$ will be the sum of the winding number at each zero point. Therefore this result obtained from Duan's $\phi$-mapping topological current theory is the same as that given in Ref. \cite{Cunhab}, while it provides a natural approach to obtain the topological charge.

Significantly, if the boundary curve $C=\partial\Sigma$ in the manifold encloses no zero point of $\phi$, we must have $Q$=0 from (\ref{qcharge}). Or if two different closed curves enclose the same zero points, the corresponding topological charges must equal. On the other hand, taking $\Sigma$ as the manifold of $x^{i}$ for certain $t$, it will give the total topological charge of black hole PSs. So $Q$ can be used to characterize different spaces.

Furthermore, following Duan's $\phi$-mapping topological current theory, we are allowed to examine the inner structure of the topological charge. Considering there are $N$ zero points of $\phi$ and the Jacobi determinant $J^{0}\left(\frac{\phi}{x}\right)\neq0$, the solution of $\phi$=0 can be expressed as
\begin{equation}
 x^i=z^{i}_{n}(t), \quad n=1,2,...,N.
\end{equation}
Near the zero points of $\phi$, $\delta^{2}(\phi)$ can be expressed as
\begin{equation}
 \delta^{2}(\phi)=\sum_{n=1}^{N}\alpha_{n}J^{0}\left(\frac{\phi}{x}\right)\bigg|_{x=z_n},
\end{equation}
where $\alpha_{n}$ is positive expanding coefficients.

According to Duan's topological current theory \cite{Duana,Duanb}, the winding number of the $n$-th zero point is expressed as $w_{n}=w(\phi, z_n)=\alpha_n J^0\left(\frac{\phi}{x}\right)\big|_{x=z_n}$. Considering $\alpha_i$ is positive, we have
\begin{equation}
 \alpha_i=\frac{|w(\phi, z_n)|}{|J^0\left(\frac{\phi}{x}\right)\big|_{x=z_n}}.
\end{equation}
The $\phi$-mapping Hopf index $\beta_i$ and the Brouwer degree $\eta_i$ at zero point $z_n$ are, respectively, given by
\begin{equation}
 \beta_n=|w(\phi, z_n)|,\quad
 \eta_n=\frac{J^{0}\left(\frac{\phi}{x}\right)}{|J^{0}\left(\frac{\phi}{x}\right)|_{x=z_n}}.
\end{equation}
Thus,
\begin{equation}
 J^{0}\left(\frac{\phi}{x}\right)\delta^2(\phi)=\sum_{n=1}^{N}\beta_n\eta_n\delta^2(x-z_n).
\end{equation}
Finally, the topological charge can be expressed as
\begin{equation}
 Q=\sum_{n=1}^{N}w_n=\sum_{n=1}^{N}\beta_n\eta_n.
\end{equation}
This relation reflects the inner structure of the topological charge.

As shown above, we suppose $J^{0}\left(\frac{\phi}{x}\right)\neq0$. However if this condition violates, there will be the phenomenon, the generation or annihilation. In order to show it, we suppose at least one component of the Jacobi tensor does not vanish, say $J^{1}\left(\frac{\phi}{x}\right)\neq0$. Therefore, according to (\ref{uxt}), we have
\begin{equation}
 \frac{dx^1}{dt}\bigg|_{(t_*,z_n)}=\frac{J^{1}\left(\frac{\phi}{x}\right)}
    {J^{0}\left(\frac{\phi}{x}\right)}\bigg|_{(t_*,z_n)}=\infty,
\end{equation}
which gives
\begin{equation}
 \frac{dt}{dx^1}|_{(t_*,z_n)}=0.
\end{equation}
Then at the critical point ($t_*$, $z_n$), we have the following Taylor expansion:
\begin{equation}
 t-t_*=\frac{1}{2}\frac{d^2t}{d(x^1)^2}\big|_{(t_*,z_n)}(x^1-z^1_n)^2.
\end{equation}
Because that the topological current is identically conserved, the topological charge of these two defects must be opposite at the critical point. When $\frac{d^2t}{d(x^1)^2}|_{(t_*,z_n)}<0$ or $>0$, it represents an annihilation or generation \cite{Fu}.

\section{Black holes and topological charge}
\label{bhatc}

As shown above, the topological charge $Q$ equals the sum of the winding number of each zero point of $\phi$ for given $\Sigma$. For a black hole, we takes $\Sigma$ as a full exterior region outside of the outer horizon. This will give us the total topological charge of the black hole PSs, and which can be used to characterize different spacetimes.

\subsection{Asymptotically flat black holes}
\label{bafbh}

\begin{figure}
\includegraphics[width=7cm]{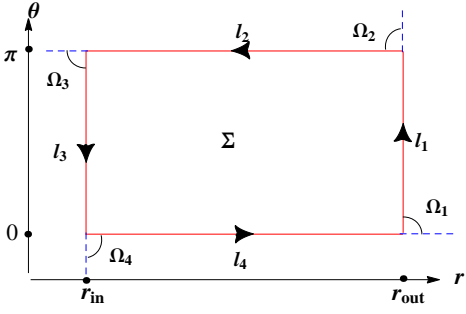}
\caption{Representation of the contour $C=\sum_i\cup l_i$ (which encloses $\Sigma$) on the
($r$, $\theta$) plane. The curve $C$ has positive orientation marked with the black arrows. $r_{\texttt{in}}$ denotes the inner boundary of $\Sigma$, which equals $r_{\texttt{h}}$ for a black hole. $r_{\texttt{out}}$ is the outer boundary, which equals $\infty$ for asymptotically flat or AdS black holes, and $r_\texttt{c}$ for dS black holes. The angle $\Omega_i$ represents the change of the direction of vector $n$ on the joint of two $l_i$.}\label{Contrth}
\end{figure}

Here we first consider an asymptotically flat black hole with the solution described by (\ref{metric}). At $r\rightarrow\infty$, the metric functions have the following asymptotic behaviors:
\begin{eqnarray}
 f(r)&\sim& 1-\frac{1}{r}+\mathcal{O}\left(\frac{1}{r^2}\right), \label{a1}\\
  g(r)&\sim& 1-\frac{1}{r}+\mathcal{O}\left(\frac{1}{r^2}\right),\\
   h(r) &\sim& r^2.\label{a2}
\end{eqnarray}
Considering the $i$-th zero point of $\phi$ is enclosed by a piece-wise smooth and positive oriented closed curve $C_i$ and other zero points are out of it, the winding number of the vector is
\begin{eqnarray}
 w_i=\frac{1}{2\pi}\oint_{C_i}d\Omega,\label{windingnumber}
\end{eqnarray}
where $\Omega=\arctan(\phi^2/\phi^1)$. Then the total charge will be
\begin{eqnarray}
 Q=\sum_iw_i.
\end{eqnarray}
On the other hand, due to the $\delta$ function in the topological current (\ref{qcharge}), the total charge of the black hole system can be calculated as
\begin{eqnarray}
 Q=\frac{1}{2\pi}\oint_{C}d\Omega.
\end{eqnarray}
Here we adopt the contour $C$ defined in Ref. \cite{Cunhab}, where $C=\sum_i\cup l_i$ or the union of four line segments $l_1\sim l_4$: $\{r=\infty, 0\leq\theta\leq\pi\}\cup\{\theta=\pi, r_{\texttt{h}}\leq r<\infty\}\cup\{r=r_{\texttt{h}}, 0\leq\theta\leq\pi\}\cup\{\theta=0, r_{\texttt{h}}\leq r<\infty\}$; see Fig. \ref{Contrth} with the black arrows denoting the direction. In order to calculate $Q$, we examine the vector field $\phi^a$ on these line segments. For simplicity, we list $\phi^a$
\begin{eqnarray}
 \phi^r&=&\frac{hf'-fh'}{2h^{\frac{3}{2}}\sin\theta}\sqrt{\frac{g}{f}},\\
 \phi^\theta&=&-\frac{\sqrt{f}\cos\theta}{h\sin^2\theta}.
\end{eqnarray}

At the horizon, one has $f(r_{\texttt{h}})=g(r_{\texttt{h}})=0$, while $\sqrt{g/f}|_{r_{\texttt{h}}}$ keeps finite. Therefore, we arrive at
\begin{eqnarray}
 \phi_{l_3}^{r}(r\rightarrow r^{+}_{\texttt{h}})>0,\quad \phi_{l_3}^{\theta}(r\rightarrow r^{+}_{\texttt{h}})\rightarrow0,
\end{eqnarray}
where $f'(r_{\texttt{h}})>0$ is used. So as shown in Fig. \ref{Contrth}, the vector $\phi$ is horizontal to the right at the horizon and thus $\Omega_{l_3}$=0. At $\theta$=0 and $\pi$, we, respectively, have
\begin{eqnarray}
& \phi^r_{l_4}(\theta\rightarrow0^+)\sim \frac{1}{\theta},\quad \phi^\theta_{l_4}(\theta\rightarrow0^+)\sim -\frac{1}{\theta^2},\\
& \phi^r_{l_2}(\theta\rightarrow\pi^-)\sim \frac{1}{\pi-\theta},\quad \phi^\theta_{l_2}(\theta\rightarrow\pi^-)\sim \frac{1}{(\pi-\theta)^2}.
\end{eqnarray}
So we have $\Omega_{l_2}=\frac{\pi}{2}$ and $\Omega_{l_4}=-\frac{\pi}{2}$. These imply that along line segments $l_{2,3,4}$, $\Omega$ does not change, so one has $\Delta\Omega_{l_2}=\Delta\Omega_{l_3}=\Delta\Omega_{l_4}=0$, and thus $\Omega_3=\Omega_4=-\frac{\pi}{2}$. Last, let us consider $\Omega$ along $l_1$. When $r\rightarrow\infty$,
\begin{eqnarray}
 \phi^r_{l_1}(r\rightarrow\infty)\propto-\frac{1}{r^2\sin\theta},\; \phi^\theta_{l_1}(r\rightarrow\infty)=-\frac{\cos\theta}{r^2\sin^2\theta}.
\end{eqnarray}
Considering that both $\phi^r_{l_1}$ and $\phi^\theta_{l_1}$ are negative, we have $\Omega_{l_1}=\pi+\arctan(\cot\theta)$. When $\theta$ varies from 0 to $\pi$, $\Omega_{l_1}$ monotonically decreases from $3\pi/2$ to $\pi/2$, or from $-\pi/2$ to $\pi/2$, which means that the vector $\phi$ changes smoothly at points ($r$, $\theta$)=($\infty$, $0$) and ($\infty$, $\pi$). So we have $\Omega_1=\Omega_2=0$ and
\begin{eqnarray}
 \Delta\Omega_{l_1}=\int_{l_1}d\Omega=-\pi.
\end{eqnarray}
Therefore, the total topological charge is
\begin{eqnarray}
 Q=\frac{1}{2\pi}\big(\Delta\Omega_{l_1}+\Delta\Omega_{l_2}
   +\Delta\Omega_{l_3}+\Delta\Omega_{l_4}\nonumber\\
   +\Omega_1+\Omega_2+\Omega_3+\Omega_4\big)=-1.
\end{eqnarray}
It is clear that our result confirms that of Ref. \cite{Cunhab} when the black hole spin is set to zero. Note that here we divide the total topological charge into eight parts. Four parts are for the curves, which are the same as  Ref. \cite{Cunhab}, while others describe the changes of the direction of the vector field at these four vertices. This division makes the calculation of $Q$ more clear.

In Ref. \cite{Cunhab}, the authors suggested that the total topological charge $Q$ may change for other boundary behaviors. We will consider this issue in the following.

\subsection{Asymptotically AdS black holes}

Here we would like to extend the study to some other asymptotical behaviors beyond the asymptotically flat case. Let us first consider an asymptotically AdS black hole case, which has the following behaviors
\begin{eqnarray}
 f(r)&\sim& \frac{r^2}{l^2}+1-\frac{1}{r}+\mathcal{O}\left(\frac{1}{r^2}\right), \label{b1} \\
  g(r)&\sim& \frac{r^2}{l^2}+1-\frac{1}{r}+\mathcal{O}\left(\frac{1}{r^2}\right),\\
   h(r) &\sim& r^2,\label{b2}
\end{eqnarray}
where $l$ is the AdS radius. Note that these asymptotical behaviors only modify the boundary condition at $r\rightarrow\infty$, so the calculation along $l_2$, $l_3$, and $l_4$ is the same as the asymptotically flat case. When $r\rightarrow\infty$, we have for a finite $l$
\begin{eqnarray}
 \phi^r_{l_1}(r\rightarrow\infty)\propto-\frac{1}{r^2\sin\theta},\; \phi^\theta_{l_1}(r\rightarrow\infty)=-\frac{\cos\theta}{lr\sin^2\theta}.
\end{eqnarray}
A detailed analysis shows that $\Omega$ keeps constant value $-\frac{\pi}{2}$ for $\theta\in$ (0, $\frac{\pi}{2}$), and constant value $\frac{\pi}{2}$ for $\theta\in$ ($\frac{\pi}{2}$, $\pi$). Therefore $\Omega$ only changes at $\theta=\frac{\pi}{2}$, which gives $\Delta\Omega_{l_1}=-\pi$. Note that the minus sign comes from the fact that the angle changes in clockwise rotation at $\theta=\frac{\pi}{2}$. Similarly, we still have $\Omega_1=\Omega_2=0$. Then combining with the value of $\Omega$ along other line segments, we obtain
\begin{eqnarray}
 Q=-1,
\end{eqnarray}
which obviously is the same as the asymptotically flat black hole case.

\subsection{Asymptotically dS black holes}

Now we turn to the asymptotically dS black hole case, which can be described by the following behaviors at large $r$
\begin{eqnarray}
 f(r)&\sim& -\frac{r^2}{l^2}+1-\frac{1}{r}+\mathcal{O}\left(\frac{1}{r^2}\right), \label{c1} \\
  g(r)&\sim& -\frac{r^2}{l^2}+1-\frac{1}{r}+\mathcal{O}\left(\frac{1}{r^2}\right),\\
   h(r) &\sim& r^2.\label{c2}
\end{eqnarray}
Different from the asymptotically flat and AdS cases, besides the black hole horizon, the spacetime admits a cosmological horizon with radius $r_\texttt{c}>r_{\texttt{h}}$, which is also a root of $g(r_\texttt{c})=f(r_\texttt{c})=0$. So for this case, we only consider the case $r_{\texttt{h}}\leq r\leq r_\texttt{c}$. Similarly, we also only need to consider the line segment $l_1$. Considering $g(r_\texttt{c})=f(r_\texttt{c})=0$, we have
\begin{eqnarray}
 \phi_{l_3}^{r}(r\rightarrow r^{-}_{\texttt{c}})<0,\quad \phi_{l_3}^{\theta}(r\rightarrow r^{-}_{\texttt{c}})=0.
\end{eqnarray}
Therefore, along $l_1$, the direction of $\phi$ is horizontal to the left, which indicates that $\Omega_{l_1}=\pi$ or $-\pi$. Therefore $\Delta\Omega_{l_1}=0$ and $\Omega_1=\Omega_2=-\frac{\pi}{2}$. Finally, we achieve
\begin{eqnarray}
 Q=\frac{1}{2\pi}\left(\Omega_1+\Omega_2+\Omega_3+\Omega_4\right)=-1,
\end{eqnarray}
where $\Delta\Omega_{l_i}=0$ ($i$=1-4) is considered.

In summary, we in this section calculate the topological charge of PS for the asymptotically flat, AdS, and dS black holes in GR. The results show that all of them have a total topological charge $Q$=-1.

Before ending this section, we would like to give several comments on this result. First, since all the black holes with different asymptotical behaviors share the same value of topological charge, they are in the same topological class from the viewpoint of the topology of the black hole PS. Second, there exists at least one unstable PS as suggested in Ref. \cite{Cunhab}. However the number of the PSs are not limited to one. Or one can say that these black holes have an odd number of PSs, while there is always one more unstable photon sphere than the stable photon sphere. Third, as indicated from above and Ref. \cite{Cunhab}, the nonrotating and rotating black hole in asymptotically flat spacetimes have the same topological charge, which indicates that the black hole spin does not affect the topological property. Thus we conjecture Kerr-AdS/dS black holes also have $Q$=-1, and thus they are in the same topological class. Moreover, if other black hole parameters do not affect the asymptotical behaviors, the topological charge stays unchanged.

\section{Dyonic black holes}
\label{dbh}

Here we would like to consider a specific case, the Dyonic black hole solution \cite{Lu}, which has a rich horizon and PS structures.

The action describing an asymptotically flat solution is
\begin{eqnarray}
 S=\frac{1}{16\pi}\int\sqrt{-g}d^4x\left(R-\alpha_1F^2-\alpha_2
   \left((F^2)^2-2F^{(4)}\right)\right),\nonumber
\end{eqnarray}
where the field strength is $F^2=-F^{\mu}_{\nu}F_{\mu}^{\nu}$ and $F^{(4)}=F^{\mu}_{\nu}F^{\nu}_{\rho}F^{\rho}_{\sigma}F^{\sigma}_{\mu}$. Solving the field equation, an exact solution of static and spherically-symmetric dyonic black hole reads \cite{Lu}
\begin{eqnarray}
 ds^2&=&-f(r)+\frac{dr^2}{f(r)}+r^2(d\theta^2+\sin^2\theta d\varphi^2),\label{mett}\\
 f(r)&=&1-\frac{2M}{r}+\frac{\alpha_1 p^2}{r^2}
       +\frac{q^2}{\alpha_{1}r^2} \;_2 F_1\left(\frac{1}{4},1,\frac{5}{4};-\frac{4p^2\alpha_2}{\alpha_1 r^4}\right).\nonumber
\end{eqnarray}
It is clear that each black hole is characterized by a set of parameters ($M$, $q$, $p$), which is associated with the mass, electric and magnetic charges, respectively. This black hole solution also satisfies the dominant energy condition.

Combining with (\ref{metric}), we see that for this black hole, $g(r)=f(r)$, $h(r)=r^2$, so our discussion in the last section holds. Then the normalized vectors are
\begin{eqnarray}
 n^{r}&=&\frac{r f'-2 f}{\sqrt{\left(r f'-2 f\right)^2+4 f
   \cot ^2(\theta )}},\\
 n^{\theta}&=&-\frac{2 \sqrt{f} \cot (\theta )}{\sqrt{\left(r f'-2
   f\right)^2+4 f \cot ^2(\theta )}}.
\end{eqnarray}
Moreover since this black hole is an asymptotically flat solution, we must have the total topological charge $Q$=-1. In the following we will perform the detailed study of the topological property of the black hole PS.

\subsection{Case one: $M=\frac{67}{10}$, $p=\sqrt{\frac{396}{443}}$, $\alpha_1=1$, $\alpha_2=\frac{196249}{1584}$, $q=6.65$}

Under this case, the black hole has two horizons located at $r_{\texttt{h1}}=0.3873$ and $r_{\texttt{h2}}=7.6249$.

\begin{figure}
\includegraphics[width=6cm]{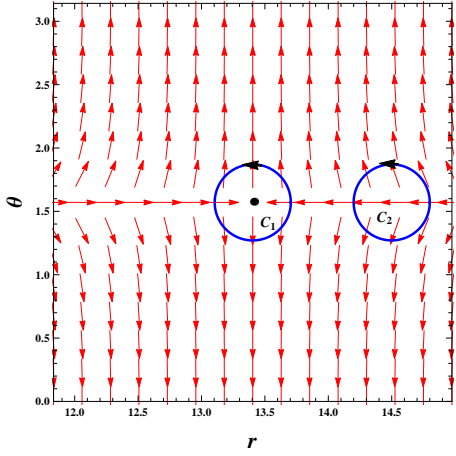}
\caption{Behavior of the normalized vector $n$ on the ($r$, $\theta$) plane for case one. The black dot denotes the zero point of $n$. $C_1$ and $C_2$ are two closed curves.}\label{pDyonCa1D}
\end{figure}

We plot the normalized vector $n$ in Fig. \ref{pDyonCa1D}. The picture is quite similar to the Schwarzschild black hole case given in \cite{Cunhab}, where only one zero point of $n$ is presented. At a first glance, one finds there exists a zero point near $r=13.5$ and $\theta=\pi/2$. Solving the equation of PS (\ref{pphoton}), we find this black hole has only one PS at $r_{\texttt{ps}}=13.4041$, which exactly coincides with the zero point of $\phi$. Since there is only one sphere photon, the topological charge or the winding number associated with this PS must be $Q=-1$. Any closed curve enclosing this PS will produce $Q=-1$, while other closed curves will give $Q=0$. The result can be obtained by observing the behavior of $n$ from Fig. \ref{pDyonCa1D}. However, we should perform the calculation of the topological charge. In the following, we attempt to calculate the topological charge (\ref{windingnumber}) for the vector along closed circular curves $C_1$ and $C_2$.

\begin{figure}
\center{\subfigure[]{\label{DomegaCL}
\includegraphics[width=6cm]{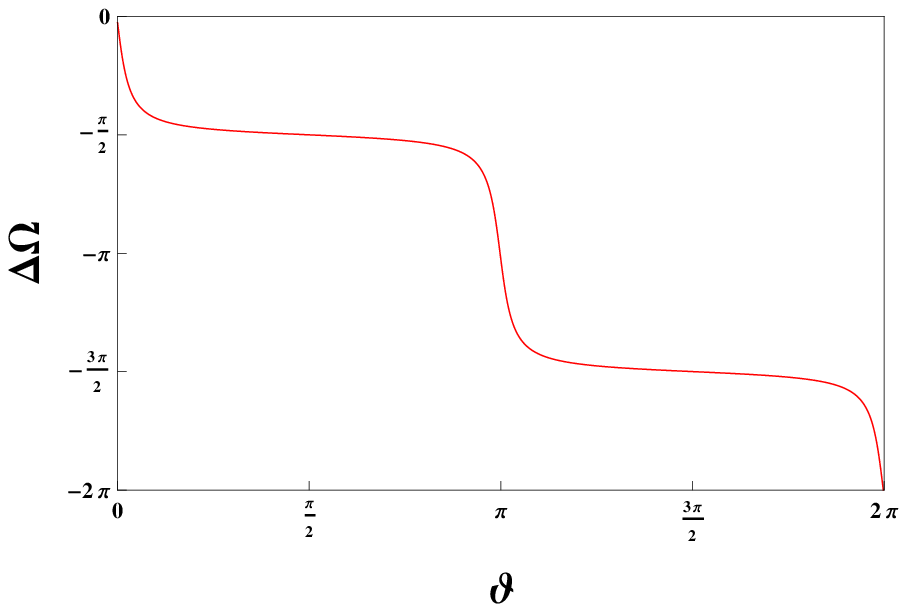}}
\subfigure[]{\label{PPDomegaC2}
\includegraphics[width=6cm]{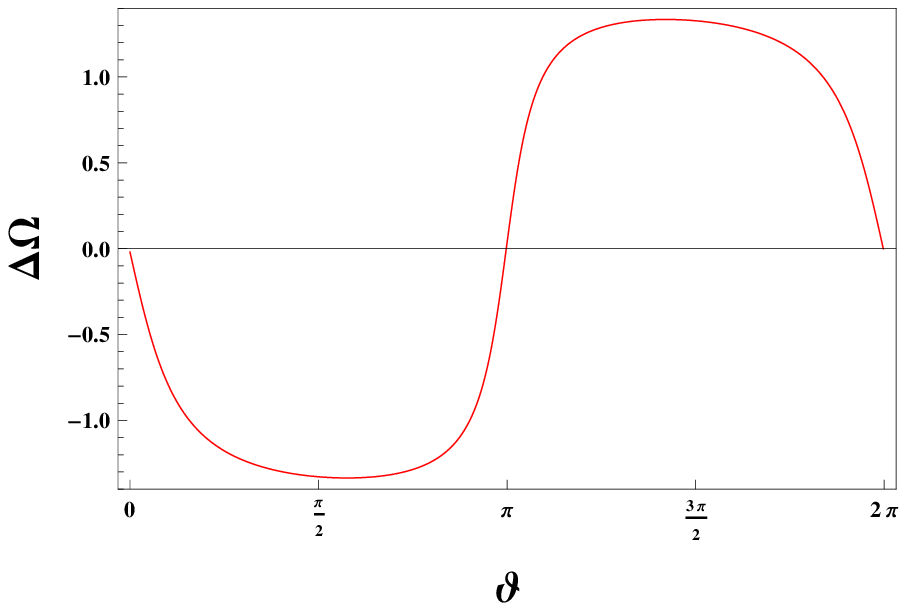}}}
\caption{$\Delta\Omega$ as a function of $\vartheta$. (a) for $C_1$, and (b) for $C_2$.}\label{ppDomegaCL}
\end{figure}

Considering that
\begin{equation}
 \Omega=\arctan\left(\frac{\phi^2}{\phi^1}\right)=\arctan\left(\frac{n^2}{n^1}\right).
\end{equation}
Then, we obtain
\begin{equation}
 d\Omega=\frac{n^1dn^2-n^2dn^1}{(n^1)^2+(n^2)^2}=\epsilon_{ab}n^adn^b.\label{shu}
\end{equation}
Therefore, the topological charge can be rewritten as
\begin{equation}
 Q=\frac{\Delta\Omega}{2\pi}=\frac{1}{2\pi}\oint_C \epsilon_{ab}n^a\partial_{i}n^bdx^{i}.
\end{equation}
Here we parametrize the closed curves $C_1$ and $C_2$ by the angle $\vartheta$ $\in$ ($0$, $2\pi$) as
\begin{eqnarray}
\left\{
\begin{aligned}
 r&=a\cos\vartheta+r_0, \\
 \theta&=b\sin\vartheta+\frac{\pi}{2}.
\end{aligned}
\right.\label{pfs}
\end{eqnarray}
We choose $(a, b, r_0)$=(0.3, 0.3, 13.4041) for $C_1$, and (0.3, 0.3, 14.5) for $C_2$. Then we calculate $\Delta\Omega$ along $C_1$ and $C_2$ by using (\ref{shu}). The result is shown in Fig. \ref{ppDomegaCL}. For $C_1$, see Fig. \ref{DomegaCL}, we find $\Delta\Omega$ decreases with $\vartheta$, and approaches -2$\pi$ at $\vartheta=2\pi$. Thus the topological charge of the vector along $C_1$ is $Q=-1$. While for $C_2$ shown in Fig. \ref{PPDomegaC2}, $\Delta\Omega$ first decreases, then increases, and finally decreases again. After a loop, $\Delta\Omega$ vanishes, and which implies that $Q=0$ for $C_2$. The difference between them is because that $C_1$ encloses a single PS, while $C_2$ does not. This also reflects the $\delta$ function encoded in the topological current (\ref{qcharge}).

This result also confirms our conclusion that the topological charge $Q$=-1 for asymptotically flat black holes. Note that in Ref. \cite{Cunhab}, the authors dubbed the PSs with $Q$=-1 or 1 as the standard or exotic PSs, respectively. Actually, we can find that the PS with $Q$=-1 is unstable, while the one with $Q$=1 is stable.

\begin{table}[h]
\setlength{\tabcolsep}{1mm}{
\begin{center}
\begin{tabular}{cccccccccc}
  \hline\hline
 &$C_1$&$C_2$&$C_3$&$C_4$&$C_5$&$C_6$&$C_8$&$C_9$&$C_{10}$\\\hline
$a$&0.3&0.3&0.5&0.5&0.5&5.7&1.0&1.0&4.0\\
$b$&0.3&0.3&0.1&0.1&0.1&0.8&0.1&0.1&0.5\\
$r_0$&13.40&14.50&2.31&6.33&12.51&9.42&7.64&11.53&9.58\\\hline\hline
\end{tabular}
\caption{Parametric coefficients of closed curves $C_i$ ($i$=1-6, 8-10).}\label{tab1}
\end{center}}
\end{table}

\begin{table}[h]
\setlength{\tabcolsep}{5mm}{
\begin{center}
\begin{tabular}{ccccc}
  \hline\hline
 &$a$&$b$&$r_0$&$\vartheta$\\\hline
 $I_1$&0.5&0.2&12.51& ($\pi$, $2\pi$)\\
 $I_2$&5.6&0.8&7.41& (0, \;$\pi$)\\
 $I_3$&0.5&0.2&2.31& ($\pi$, $2\pi$)\\
 $I_4$&4.6&0.6&7.41& ($\pi$, \;0)\\\hline\hline
\end{tabular}
\caption{Parametric coefficients of closed curves $C_7$.}\label{tab2}
\end{center}}
\end{table}

\subsection{Case two: $M=\frac{67}{10}$, $p=\sqrt{\frac{396}{443}}$, $\alpha_1=1$, $\alpha_2=\frac{196249}{1584}$, $q=6.85$}

Generally, a black hole possesses one PS. However, for this kind of black hole, in some parameter regions, it can have more than one PS \cite{Cunhab}. This provides us a good opportunity to study the topological properties of black hole PS when its number is larger than one.

Here we focus on case two: $M=\frac{67}{10}$, $p=\sqrt{\frac{396}{443}}$, $\alpha_1=1$, $\alpha_2=\frac{196249}{1584}$, $q=6.85$. Solving $f(r_{\texttt{h}})$=0, we find that the black hole has two horizons located at $r_{\texttt{h}}$=0.6302 and 1.6643, respectively. Solving the PS equation, we observe three PSs at $r_{\texttt{ps1}}$=2.3066, $r_{\texttt{ps2}}$=6.3340, and $r_{\texttt{ps3}}$=12.5075.

\begin{figure}
\center{\subfigure[]{\label{DCas2a}
\includegraphics[width=5.5cm]{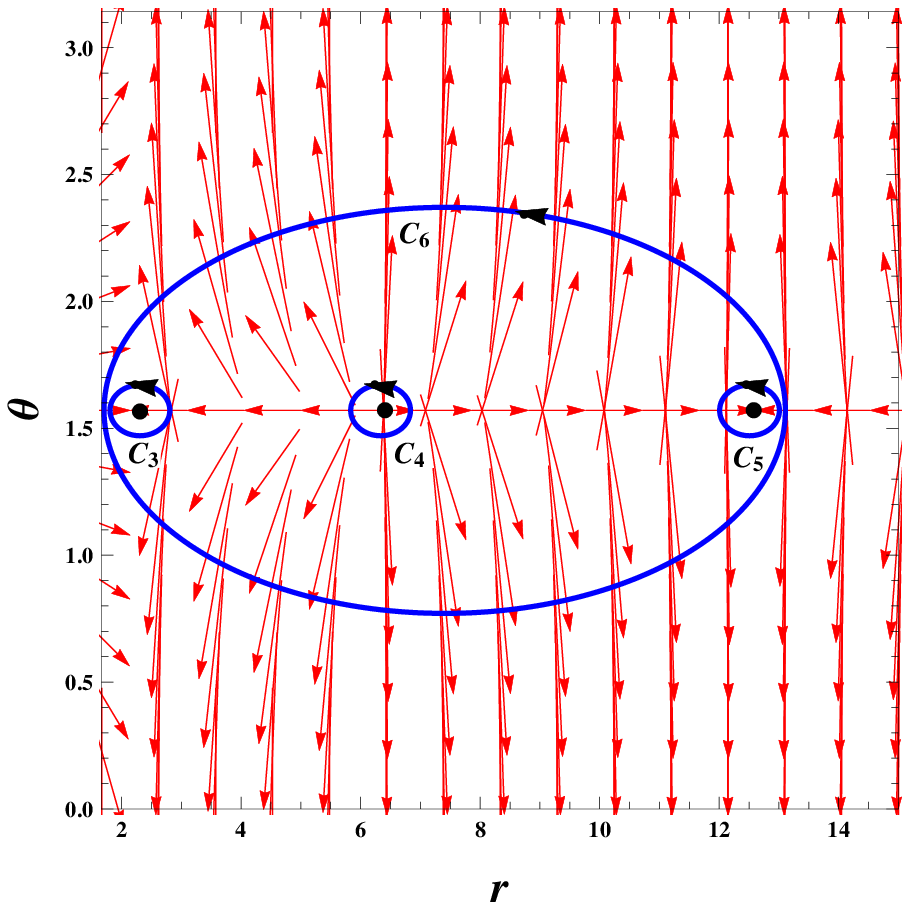}}
\subfigure[]{\label{PPCas2b}
\includegraphics[width=5.5cm]{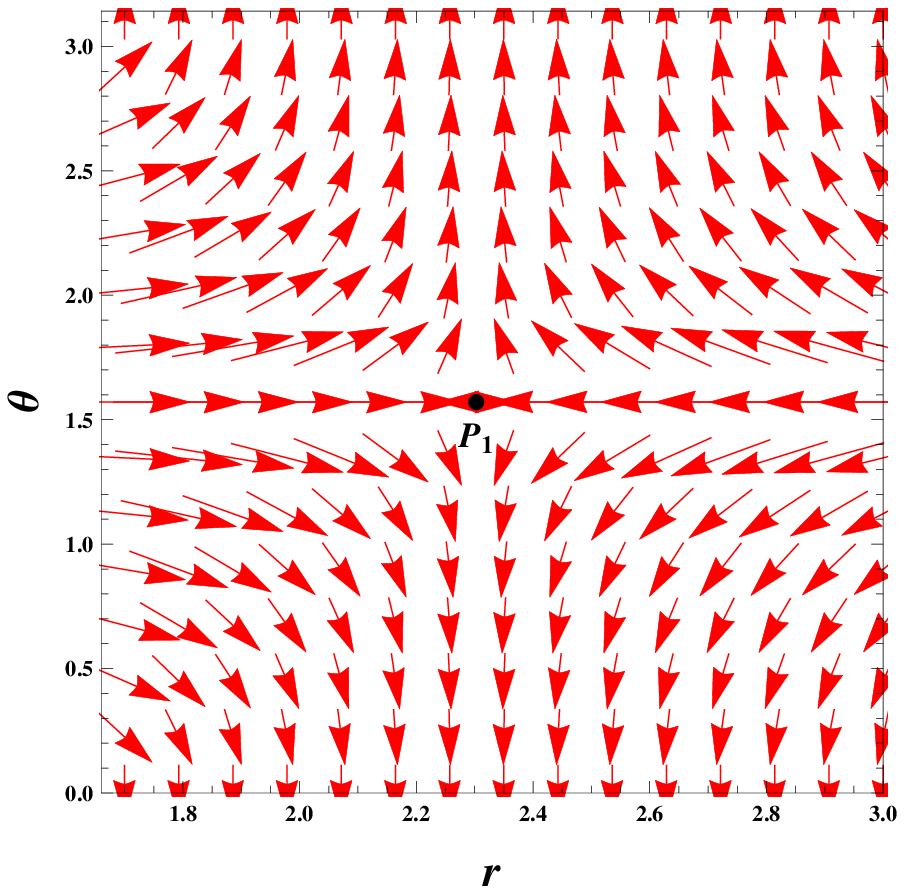}}
\subfigure[]{\label{DoCas2c}
\includegraphics[width=5.5cm]{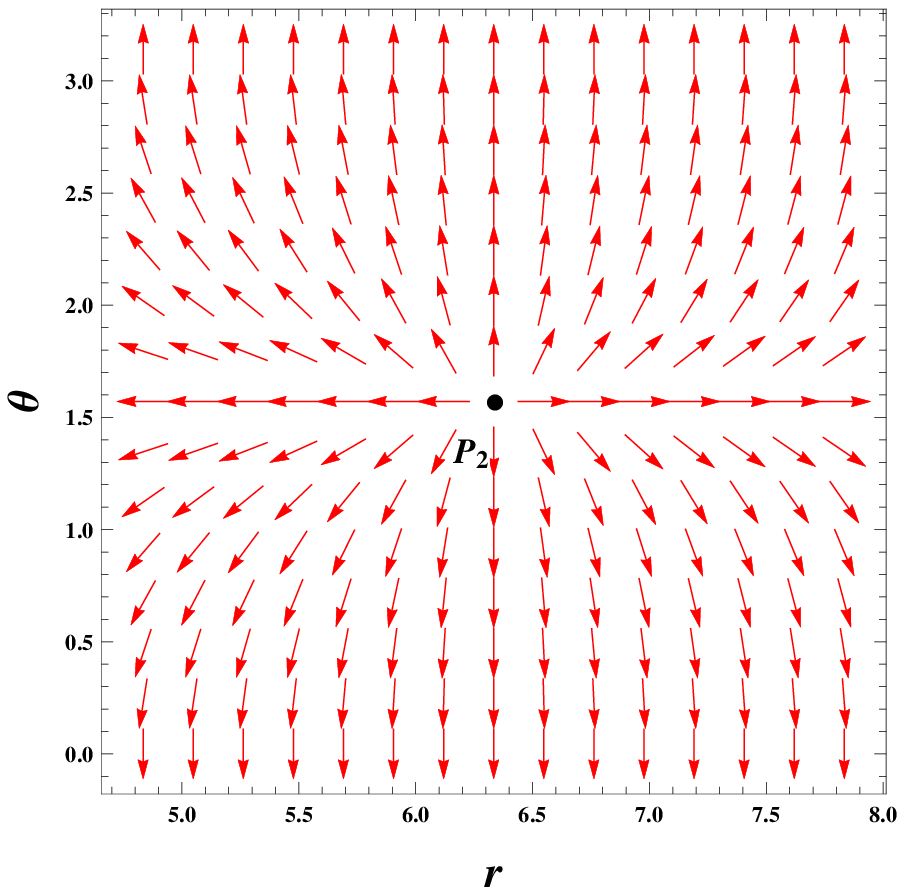}}
\subfigure[]{\label{PPCas2d}
\includegraphics[width=5.5cm]{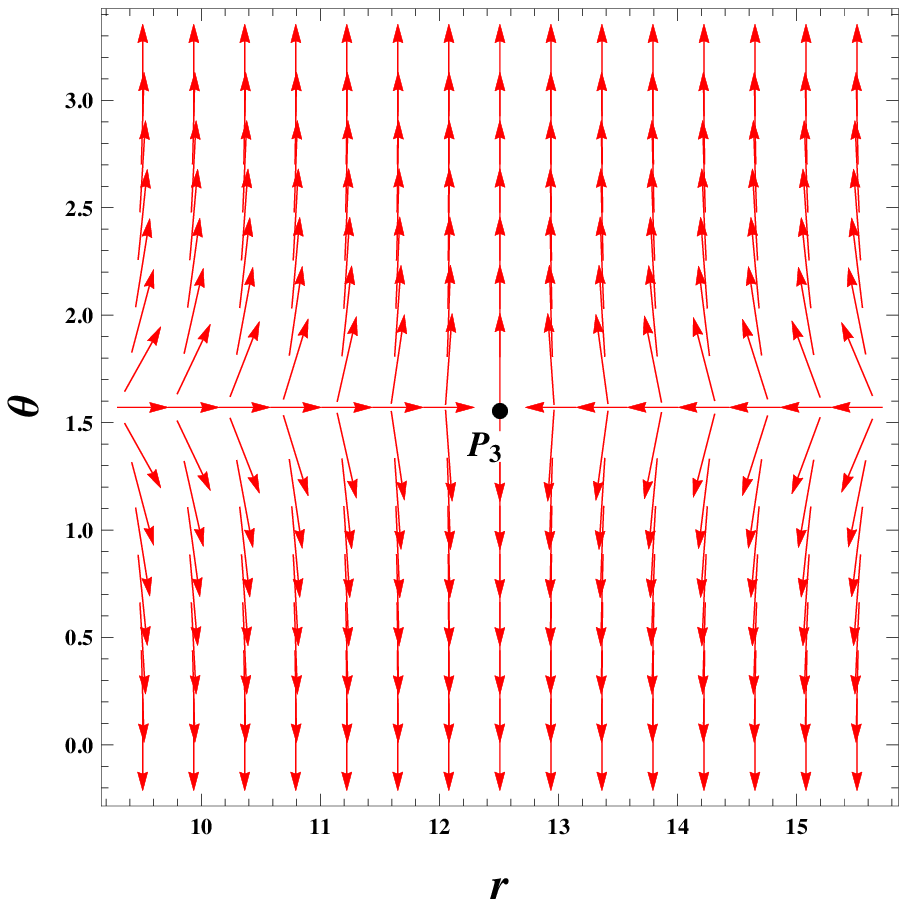}}}
\caption{Behavior of the normalized vector $n$ on the ($r$, $\theta$) plane for case two. Black dots denote the zero points of $n$. (a) Three zero points are completely shown. (b) Behavior of $n$ near zero point $P_1$. (c) Behavior of $n$ near zero point $P_2$. (d) Behavior of $n$ near zero point $P_3$.}\label{ppDoCas2d}
\end{figure}

In order to clearly display the behavior of vector $n$, we describe it on the ($r$, $\theta$) plane in Fig. \ref{ppDoCas2d}. Obviously, such behavior is significantly different from the one given in Fig. \ref{pDyonCa1D}. We observe that there are three zero points of $n$, which exactly coincide with the locations of PSs. So the PSs can be treated as topological defects of the black hole system even when more PSs are included. We show the local behaviors of $n$ near these three zero points in Figs. \ref{PPCas2b}, \ref{DoCas2c}, and \ref{PPCas2d}. Obviously, the vector $n$ has similar behaviors around points $P_1$ and $P_3$, while near $P_2$, it like the electric field of a positive charge. So we conjecture that the topological charges associated with $P_1$ and $P_{3}$ are the same, while different from that of $P_2$.

\begin{figure}
\includegraphics[width=6cm]{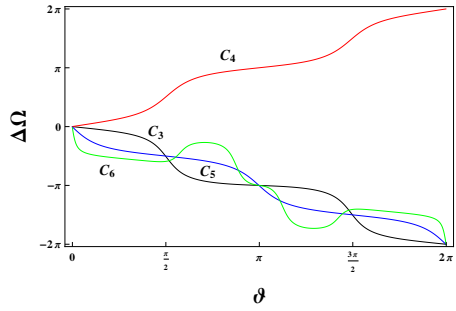}
\caption{$\Delta\Omega$ as a function of $\vartheta$ for $C_3$, $C_4$, $C_5$, and $C_6$.}\label{pWindingN5}
\end{figure}

To examine the topological charges associated with these three PSs, we construct three closed curves enclosed by $C_{3,4,5}$ shown in Fig. \ref{DCas2a}, respectively. Moreover, we also construct a large closed curve $C_6$, which contains these three PSs. These parametrized forms share the same expression given in Eq. (\ref{pfs}) while with different coefficients given in Table \ref{tab1}. Then we numerically calculate $\Delta\Omega$ along these curves. The results are given in Fig. \ref{pWindingN5}. With the increase of $\vartheta$, $\Delta\Omega$ decreases along $C_3$ and $C_5$, and increases along $C_4$. For complete closed curves, we easily observe that the topological charge $Q$=-1 for $P_1$ and $P_3$, and $Q$=1 for $P_2$. This confirms our conjecture that $P_1$ and $P_3$ have the same topological charge, while it is different for $P_2$. So we can say that the PSs at $P_1$ and $P_3$ are standard, while the one at $P_2$ is exotic.

On the other hand, since the black hole has three PSs and the total topological charge for its PSs is minus one, any closed curves enclosing these three points must produce $Q$=-1. To check this result, we calculate $\Delta\Omega$ along $C_6$. Nevertheless, $\Delta\Omega$ exhibits a nonmonotonic behavior with $\vartheta$, it gives $\Delta\Omega=-2\pi$ after a loop, which indicates $Q$=-1. Therefore, it is consistent with our analysis in Sec. \ref{bafbh} for the asymptotically flat black holes.

\begin{figure}
\center{\subfigure[]{\label{DoVect}
\includegraphics[width=6cm]{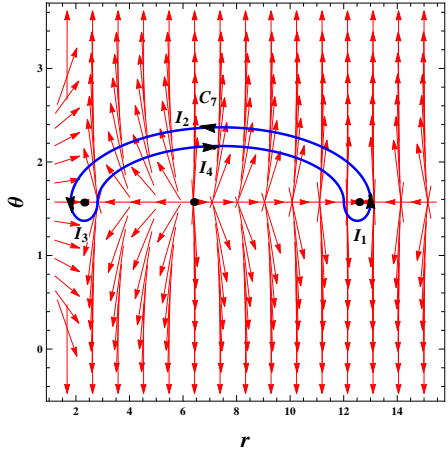}}
\subfigure[]{\label{POme}
\includegraphics[width=6cm]{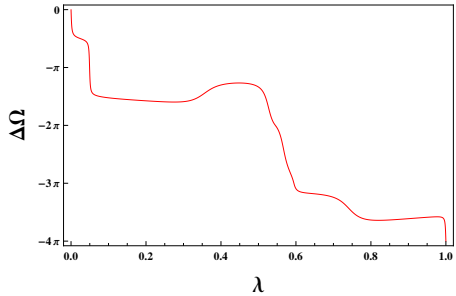}}}
\caption{(a) Schematic diagram of closed curve $C_7$. (b) $\Delta\Omega$ as a function of length parameter $\lambda$.}\label{ppOme}
\end{figure}

Interestingly, we can also construct another closed curve, see $C_7$ shown in Fig. \ref{DoVect}, which only encloses points $P_1$ and $P_3$. The curve $C_7$ is constructed by four segments, $I_1$-$I_4$, which are also parametrized as (\ref{pfs}) with the coefficients given in Table \ref{tab2}. Here we show $\Delta\Omega$ as $C_7$'s length parameter rather $\vartheta$ in Fig. \ref{POme}. Circulating the contour $C_7$ anti-clockwise $\Delta\Omega$ approaches -4$\pi$. Hence, we have the topological charge $Q$=-2 for $C_7$, which is just the sum of the charges of $C_3$ and $C_5$ as expected.

\subsection{Case three: $M=\frac{67}{10}$, $p=\sqrt{\frac{396}{443}}$, $\alpha_1=1$, $\alpha_2=\frac{196249}{1584}$, $q=7$}

As shown above, the topological charge $Q$=-1 for the dyonic black holes in cases one and two. So these black hole cases are in the same topological class, regardless of the number of PS. Here we wonder whether there exists different structure of PS from the topology. For this purpose, we focus on case three with charge $q$=7. It is easy to find that the spacetime in this case will not admit a black hole but a naked singularity. The disappearance of the horizon will make the vector flow towards $r=0$. Considering that the spacetime is singular at $r$=0, we exclude it. Thus, this solution is quite similar to the horizonless
ultra-compact object, and we conjecture $Q$=0 for the naked singularity.

\begin{figure}
\center{\subfigure[]{\label{DVectorc}
\includegraphics[width=6cm]{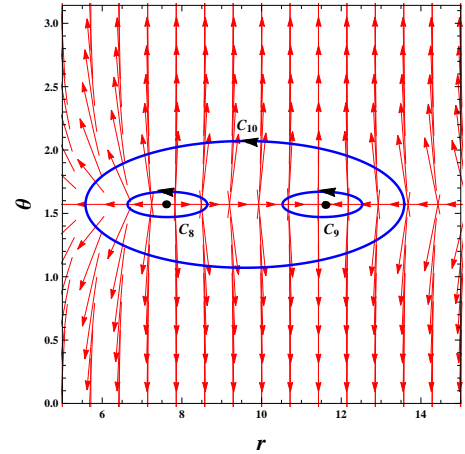}}
\subfigure[]{\label{PDeltaC}
\includegraphics[width=6cm]{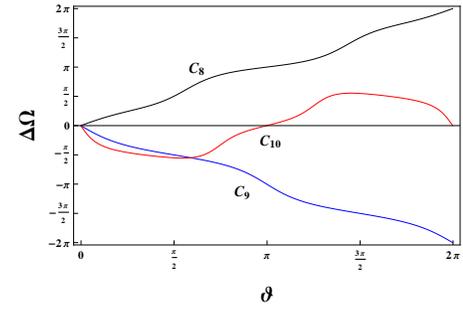}}}
\caption{(a) Behavior of the normalized vector $n$ on the ($r$, $\theta$) plane for case three. (b) $\Delta\Omega$ as a function of $\vartheta$ for $C_8$, $C_8$, and $C_{10}$.}\label{ppDeltaC}
\end{figure}

We first show the vector $n$ on the ($r$, $\theta$) plane in Fig. \ref{DVectorc}. From it, two zero points of $n$ are observed. The left one is similar to the point $P_2$, and the right one to $P_3$. Proceeding as above, we construct the closed curves $C_8$ and $C_9$, which, respectively, enclose these two zero points. Another closed curve $C_{10}$ encloses both these two points. The parametrized coefficients are given in Table \ref{tab1}. Then we calculate $\Delta\Omega$ as a function of $\vartheta$. The result is listed in Fig. \ref{PDeltaC}. It is easy to read out from the figure that $Q$=1 for $C_8$ and $Q$=-1 for $C_9$. Moreover, the topological charge vanishes for $C_{10}$. Therefore, we confirm that the naked singularity has $Q$=0 as expected. Comparing with Ref. \cite{Cunhab}, we find that this case is similar to the rotating boson star without event horizon. And thus they belong to the same topological class.

Further calculation shows that these two points approach each other with the increase of charge $q$. When $q\approx7.1$, these two points meet. Beyond that value, the spacetime will not exhibit the PS. So we always have $Q$=0.

Moreover, it is worth pointing out that when $q$=6.92, the naked singularity has four PSs located at $r_{\texttt{ps}}$=1.1358, 1.5794, 6.9090, and 12.1041 with the topological charge 1, -1, 1, -1. And the total topological charge still equals zero.

\section{Photon spheres annihilation}
\label{psa}

From the topology, two defects of opposite topological charge can annihilate, or generate from vacuum with time. For an equilibrium black hole, its PS will be determined and thus no the phenomenon of the generation or annihilation exists. However, when considering the Hawking radiation, the black hole gradually losses its mass and other charges, and the PS will be changed. Here we imagine a black hole system that absorbs the charge from its surroundings through a quasistatic process. With the continuous increase of the charge, the black hole will turn to a naked singularity. In this case, the electric charge plays the role of evolution time, so we adopt $x^0=q$. In the following, we would like to examine the change of the PS.

\begin{figure}
\includegraphics[width=6cm]{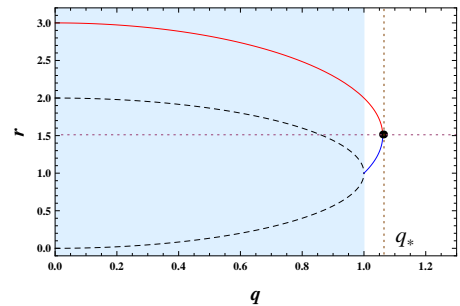}
\caption{Annihilation of two PSs (red and blue thin lines) with opposite sign of winding number. Top one has $w=$-1, and bottom one has $w$=1. The dashed lines are for two horizons. The shadow region of $q<1$ is for the black hole case, while others is for naked singularity. The parameter is $q_*=\frac{3}{2\sqrt{2}}$. The black hole mass is set to $M$=1.}\label{pWindingN}
\end{figure}

For simplicity, we set $\alpha_2$=0. Then the black hole solution in (\ref{mett}) actually describes a charged Reissner-Nordstr\"om black hole. The black hole horizons are located at $r_{\texttt{h}1,2}=M\pm\sqrt{M^2-q^2}$, and the PSs are at $r^{q}_{\texttt{ps}1,2}=\frac{1}{2}(3 M\pm\sqrt{9 M^2-8 q^2})$. For the black hole case $M>q$, we always have $r^{q}_{\texttt{ps}2}<r_{\texttt{h}1}$, which means that the inner PS is covered by the outer horizon. Thus only one PS is allowed, which is a standard one and has topological charge $Q$=-1. When the black hole is over charged $M<q$, these two horizons disappear. Then the second PS of positive topological charge $Q$=1 at $r^{q}_{\texttt{ps}2}$ takes action, which leads to $Q=0$ for the naked singularity. Thus from a black hole to a naked singularity, spacetime undergoes a topological phase transition. We describe the case in Fig. \ref{pWindingN}.

For naked singularity, with the increase of the charge $q$, these two PSs approach and annihilate at $q=q_{*}=3M/(2\sqrt{2})$. From the viewpoint of topology, this phenomenon denotes an annihilation of two PSs. Meanwhile the total topological charge does not change.

Further, expanding the charge near $q_*$, we have
\begin{eqnarray}
 &q-q_*=\frac{1}{2}q''(r-r^{q}_{\texttt{ps}})^2,\\
 &q''=\frac{d^2q}{dr^2}\big|_{(q_*, r=r^{q}_{\texttt{ps}})}=-\frac{\sqrt{2} \left(r-\frac{3 M}{2}\right)^2}{3M^2}.
\end{eqnarray}
Interestingly, one observes a negative $q''$. So this annihilation process is in accord with that froms the topological current $\phi$-mapping theory discussed in \cite{Fu}.

On the other hand, if taking the decrease of $q$ as the evolution direction, this pattern will become the generation of PSs.

\section{Discussions and conclusions}
\label{Conclusion}

In this paper, we studied the topological property of a PS for static, spherically symmetric black hole with different asymptotical behaviors in GR.

Following Duan's topological current $\phi$-mapping theory, we introduced the superpotential $V^{\mu\nu}$ and topological current $j^{\mu}$ defined on the ($r$, $\theta$) plane outside of a black hole. After some calculations, we expressed the topological charge $Q$ of PSs in terms of a $\delta$ function, see (\ref{qcharge}), which indicates that the black hole PS's topological property is closely dependent of the zero points of vector field $\phi$ enclosed by the considering parameter region. The inner structure of the topological charge is also investigated.

Then we computed the topological charge for a general black hole solution with asymptotically flat, AdS, and dS boundaries. Here we summarize two universal properties: i) $\Omega_{3}=\Omega_{4}=-\frac{\pi}{2}$, and $\Delta\Omega_{l_2}=\Delta\Omega_{l_3}=\Delta\Omega_{l_4}$=0. ii) $\Omega_1+\Omega_2+\Delta\Omega_{l_3}=-\pi$. Then after carrying out the detailed analysis, we find that all these black hole systems have the same topological charge $Q=$-1. This suggests that these black holes have at least one standard PS, the same as that for the stationary, axi-symmetric, asymptotically
flat black hole \cite{Cunhab}. If there are several PSs, the number of standard PSs should be one more than the exotic ones. Considering that the black hole spin might not change the total topological charge indicated in Ref. \cite{Cunhab}, we conjectured that the rotating black holes with asymptotically flat, AdS, and dS boundaries also admit $Q=$-1. This issue is worth for further pursuing.

As a specific example, we computed the topological charge for the dyonic black holes with $M=\frac{67}{10}$, $p=\sqrt{\frac{396}{443}}$, $\alpha_1=1$, $\alpha_2=\frac{196249}{1584}$, while with different electric charge $q$. For the case one, $q$=6.65, the black hole exhibits only one standard PS. We clearly showed that for the closed curve enclosing the PS, one has $Q$=-1. While for other closed curves, it produces $Q$=0. This reflects the property of the $\delta$ function contained in the topological current (\ref{qcharge}). For the second case, the electric charge is set to $q$=6.85. The black hole admits more than one PS, and three of them are observed. Two are standard ones associated with $Q$=-1 and one is exotic one with $Q$=1. The total topological charge still keeps -1. Therefore case two and case one are in the same topological class. Case three with $q$=7 corresponds to a naked singularity rather a black hole. Its topological charge is found to be $Q$=0, which implies that the standard and exotic PSs come in pairs. For example, when $q$=7, we observed a pair of standard and exotic PSs. However two pairs will present when $q$=6.92. The different total topological charges of black holes and naked singularities also indicate that they are in different topological classes.

Finally, we took the charge $q$ as an evolution parameter and investigated the annihilation of the standard and exotic PSs. The result is also in accord with the Duan's topological current $\phi$-mapping theory.

Our result confirms that static, spherically symmetric black holes with different asymptotically flat, AdS, and dS behaviors, respectively described by (\ref{a1})-(\ref{a2}), (\ref{b1})-(\ref{b2}), (\ref{c1})-(\ref{c2}) have at least one standard PS from the viewpoint of topology. This topological approach provides us a novel insight into the black hole PS structure. Topological phase transition can also be revealed by the charge $Q$. So we expect the topology of a black hole PS can play a novel role in investigating the astronomical phenomena related with the black hole PS. Furthermore, it is also interesting to consider other black hole or naked singularity backgrounds in a spacetime with different asymptotics.

\section*{Acknowledgements}
Shao-Wen Wei would like to thank Profs. Yu-Xiao Liu, Jie Yang, Hai-Shan Liu, Songbai Chen, and Li Zhao for helpful discussions. This work was supported by the National Natural Science Foundation of China (Grant No. 11675064) and the Fundamental Research Funds for the Central Universities (Grants No. lzujbky-2019-it21).

\end{document}